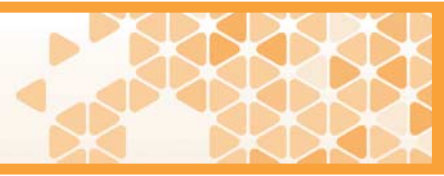

# The evolving perception of controversial movies

Luca Amendola[1], Valerio Marra[2] and Miguel Quartin[3]

**ABSTRACT** Polarization of opinion is an important feature of public debate on political, social and cultural topics. The availability of large internet databases of users' ratings has permitted quantitative analysis of polarization trends—for instance, previous studies have included analyses of controversial topics on Wikipedia, as well as the relationship between online reviews and a product's perceived quality. Here, we study the dynamics of polarization in the movie ratings collected by the Internet Movie database (IMDb) website in relation to films produced over the period 1915–2015. We define two statistical indexes, dubbed hard and soft controversiality, which quantify polarized and uniform rating distributions, respectively. We find that controversy decreases with popularity and that hard controversy is relatively rare. Our findings also suggest that more recent movies are more controversial than older ones and we detect a trend of "convergence to the mainstream" with a time scale of roughly 40–50 years. This phenomenon appears qualitatively different from trends observed in both online reviews of commercial products and in political debate, and we speculate that it may be connected with the absence of long-lived "echo chambers" in the cultural domain. This hypothesis can and should be tested by extending our analysis to other forms of cultural expression and/or to databases with different demographic user bases.

[1] Institut für Theoretische Physik, Universität Heidelberg, Heidelberg, Germany [2] Departamento de Física, Universidade Federal do Espírito Santo, Vitória, Brazil [3] Instituto de Física, Universidade Federal do Rio de Janeiro, Rio de Janeiro, Brazil Correspondence: (email: valerio.marra@me.com)

## Introduction

Polarization in public opinion has been the subject of many analyses in the past, due to its important implications in politics, social sciences, economy and marketing. The availability of large digital datasets has allowed in the last decade a quantitative real-world analysis that produced many new insights into how polarization begins and evolves, as well as on how cultural or demographic differences affect its phenomenology. This new area of research has also been accompanied by sophisticated mathematical models that simulate the exchange of opinions and reveal the statistical mechanisms that underlie opinion dynamics, as for instance in Castellano *et al.* (2009).

Most analyses have so far been devoted to polarization in regard to political themes, as a way of better understanding the origin and trends of possible conflicts among the citizens. Another area of research, of particular interest to the field of marketing, has developed around polarization of consumers' opinions about commercial items, from restaurants to home goods. Automated searches and characterizations of controversial topics, such as those raised on social media, could allow politicians or marketers to effectively identify and address complaints or concerns regarding policies or products.

A less substantial body of research has been directed towards the quantitative assessment of polarizations in taste in relation to cultural products (such as films and literature) and their evolution with time. While the question of whether modern democratic societies are becoming more or less polarized on political issues, such as gun control or abortion, has been repeatedly addressed and hotly debated (DiMaggio *et al.*, 1996; McCarty *et al.*, 2006; Baldassarri and Bearman, 2007; Fiorina and Abrams, 2008; Fischer and Mattson, 2009; Garcia *et al.*, 2015; Koutra *et al.*, 2015), much less effort has been devoted to similar questions about cultural expression, as we discuss below. Controversiality can arise as a consequence of artistic innovation, or because a work of art explores sensitive topics such as religion, politics, ethics or even simply because it appeals to a particular demographic section of society and not others. Kostelanetz (2000) claims in *A Dictionary of the Avant-Gardes*: "My basic measures of avant-garde work are esthetic innovation and initial unacceptability". A high degree of controversy can then be seen as a sign of innovative work of art, although of course it is by no means a sufficient condition.

This paper presents a contribution to this particular area by studying the evolution of controversy in the public perception of movies and its dependence on demographic factors. We employ the data collected by the Internet Movie database (IMDb) website, where viewers rate movies from 1 (worst) to 10 (best). The IMDb database is a large and continuously updated catalogue that also includes demographic details related to age, gender and geographic origin. Using the data available we are able to formulate a mathematical rationalization of trends and, at the same time, offer perspectives and terminology that could inform further future research. We expect that our methodology could be applied to other forms of expression, such as music and literature, and that the results of our enquiry could provide a suitable basis for psychological, sociological and philosophical evaluations of the issues at hand.

**Terminology**. In this study we identify two different kinds of controversiality: one in which the debate is polarized at the two extremes of the spectrum, and another in which opinions distribute evenly across it. Here, we will use the term "controversy" to refer to both these possibilities, and the term "polarization" only for the first kind of controversiality.

## Literature review

The issue of controversy in the public debate has been most often analysed on an episodic basis. A casual search of the keyword "controversy" or related terminology in any journal database retrieves hundreds of papers dealing with particular controversies in any field of sciences or humanities. Some attempts at identifying common patterns in controversies arising in science or philosophy have been presented in the literature, notably starting with the famous work on the scientific revolutions by Kuhn (1962). Although limited to analyses of a few cases in specific fields, nevertheless these works have supported the idea that controversies might have universal features.

The advent of the internet, of digital databases and of social media, has allowed researchers to analyse for the first time huge quantities of data related to all fields of human expression. Social media like Twitter and Wikipedia offer a fascinating opportunity for extensive research on automated identification of controversial topics, on their classification and on their common features. Akoglu (2014); Mejova *et al.* (2014) analyse US online news outlets and political databases in search of reliable indicators of controversiality and of language features. Yasseri *et al.* (2014) investigates the multicultural aspects of Wikipedia by searching for the most controversial topics, identified as those that were subjected to "edit wars" between editors with different views. In Garimella *et al.* (2015) the authors focus on measuring the degree of controversy in Twitter conversations on particular topics, such as news items, with the aim of automatizing the task of finding and comparing controversial news. In Koutra *et al.* (2015) the controversial topic of gun control was selected and studied across web-sites in order to track and analyse reactions to a shocking news event. Additional studies on Twitter and Wikipedia controversies include those by Conover *et al.* (2011), Yardi and Boyd (2010), Rad and Barbosa (2012) and Pennacchiotti and Popescu (2010).

All these works deal with social media and, mostly, with political controversies or other topics related to news or issues of popular interest, such as sport. Clearly such opinions are heavily influenced by demography and by events like elections, sport results, conflicts or economic developments (McCarty *et al.*, 2006; Fiorina and Abrams, 2008; Yardi and Boyd, 2010; Garcia *et al.*, 2015). One expects controversies in art and, specifically, in movies to be less affected by the latter short-term events and, in particular, to feature different temporal trends. Controversy in cultural debate has been the subject of only limited quantitative studies. In Hu *et al.* (2009), the average distribution of ratings of books and DVDs on Amazon was found to follow a universal J-shaped distribution. The same pattern is implicit in the findings of Cai *et al.* (2013) among raters of musical items. Similar research in an online collaborative platform for scholarly projects instead found a rating distribution peaked at around 75% of the maximum score (Bell and Ippolito, 2011). While the study of the average distribution of ratings is not our prime focus, here we compare our results to these findings. Godes and Silva (2012) finds a decrease of online ratings for books after initial reviews, attributed to different types of users who post reviews at different stages: fans first, casual readers later on; they report that the decrease seems to stabilize roughly one year after the first reviews. In Sun (2012a), the average and variance of book ratings are correlated to sales, and a positive correlation of variance-sales is observed when the average is below a certain threshold, perhaps because the rating scatter attracts the attention of potential readers.

Ratings have also been analysed in other contexts. Zhang *et al.* (2014) considered the temporal evolution of the average and the variance of online restaurant ratings over a period of 10 years. They find that while the average rating steadily increases over time, the variance correspondingly decreases. In Moe and Schweidel (2012) the authors find that frequent raters of home products tend to post worse and more varied ratings than casual raters, thereby increasing overall variance over time. These trends of mean and variances bear some similarities, but also important differences, with respect to our findings, as we will discuss in the next sections.

A few works performed a statistical study of film ratings, although not connected to controversiality. In what can be considered a pioneer work on social media ante litteram, Wanderer (1970) tests whether film ratings assigned by professional critics agree with those assigned by normal viewers, finding only marginal differences. Holbrook (2005) employs IMDb data to assess the same question, finding a similarly high correlation among ratings, but also detecting low correlation among movies recommended by professional critics and movie popularity (that is, frequently reviewed by ordinary users). In Moon et al. (2010), the authors find significant correlations between film ratings in internet databases, both from ordinary viewers and from professional critics, and box office revenues. Liu (2006) finds that movie audiences are more critical just after a movie is released, and that online activity correlates well with box-office revenues.

The database we employ in the current work, IMDb, has been previously analysed from different perspectives. Kostakos (2009) compares IMDb with other sources of ratings and discusses the rating bias induced by the website design. Hoßfeld et al. (2011) analysed 2 million movie ratings and argued that substantial information could be gathered by studying the standard deviations of these ratings (in contrast to just the means). Koh et al. (2010) finds that online ratings in IMDb and in a similar Chinese website, douban.com, represent the underlying perceived quality of movies in a way that differs among cultures (United States, China and Singapore). US citizens are more likely to underreport and, therefore, more likely to produce an average score that is different from the perceived average. US citizens also seem more likely to give ratings that are closer to the minimum or maximum vote than previous ones, while Chinese users do the opposite. They find their result to be in agreement with expectations based on the "individualism-collectivism" and "long-term orientation" axes of Hofstede's cultural dimensions theory (Hofstede, 2001). Otterbacher (2013) finds a clear linguistic difference in film reviews written by males and females and also on their degree of perceived usefulness to other users. To conclude, two main issues investigated in the present study, controversiality in the movie rating distributions and the long-term time trends of ratings assigned to the same movies, appear not to have been investigated in the past.

## Methods

It is not obvious how to characterize controversiality on a quantitative basis. Here, we propose two basic ways. One is when a movie gets many very positive and many very negative ratings—a so-called "love-hate movie". We term this "hard" controversiality. The other occurs when a movie splits opinions across a broader spectrum, generating discord among voters. In the extreme case, this leads to a roughly equal distribution of votes across all possible ratings. We term this phenomenon "soft" controversiality.[1]

We define the indexes $H$ and $S$ as normalized measures of hard and soft controversiality, respectively: $H$ is unity only for a completely polarized rating distribution of 1s and 10s, while $S$ is unity only when each rating is assigned by 1/10 of the users. Although both $H$ and $S$ vanish for unanimous distributions, we identify least controversial movies with movies with lowest $H$. Indeed, as discussed below, $H$ is built upon the concept of variance, and unanimity is connected with the state of lowest variance. Low $H$, high $S$ and high $H$ characterize the three broad classes into which distributions defined in a small range of values (ratings 1–10) will typically fall: peaked (i.e. a single, prominent peak), flat or polarized (i.e. two well separated peaks), respectively.

Mathematically, we define $H$ as a normalized standard deviation:

$$H = \frac{1}{c_H}\left[\sum_{i=1}^{10} p_i (r_i - \bar{r})^2\right]^{1/2} \tag{1}$$

where $p_i = v_i/N$, $v_i$ is the number of votes of the rating $r_i = i$, $N$ is the total number of votes,

$$\bar{r} = \sum_i p_i r_i \tag{2}$$

is the average rating, and $c_H^2$ is the highest possible variance. The largest value $c_H = 4.5$ is obtained if half the ratings are 1 and half are 10 (a completely polarized distribution). The use of the variance to characterize polarization has been advocated in the past by several authors (DiMaggio et al., 1996; Baldassarri and Bearman, 2007; Hu et al., 2009; Hoßfeld et al., 2011; ; Zhang et al., 2014). Note that, although high $H$ can indicate bimodality, that is, two peaks not necessarily located at the boundaries of the distribution, $H$ is not optimized for this role (for a comparison with bimodality estimators see Supplementary Information: comparison with estimators of bimodality).

Soft controversiality $S$ in turn can be defined as

$$S = 1 - \frac{1}{c_S}\left[\sum_{i=1}^{10} (p_i - 0.1)^2\right]^{1/2} \tag{3}$$

which is a square root of a $\chi^2$ statistics relative to the flat distribution with $p_i = 0.1$, normalized with $c_S = \sqrt{0.9}$ in such a way that it vanishes if all votes are given to a single rating $r_i$. To our knowledge, no estimator similar to $S$ has been introduced before in the context of rating distributions.

## Results

We evaluate then $H$ and $S$ for each of the feature movies listed in IMDb, from 1915 to 2014 (our data was collected in January 2015), with a breakdown in a few demographic categories. To avoid fluctuations arising from small-number statistics, we consider only movies with at least $N = 1000$ ratings.[2] There are 19,017 feature movies with this many ratings, with almost 440 million individual ratings. This represents, for instance, a factor of four more ratings than the data used in Kostakos (2009) and two orders of magnitude larger than in Hoßfeld et al. (2011). Figure 1 shows the number of movies as a function of the release year, while the average rating distribution for all movies and various demographic subgroups is given in Fig. 2 (for a comparison of this distribution with other online review distributions see Supplementary Information: the average rating distribution). The average rating is approximately 7, independent of the demographic subgroup considered.

We show in Fig. 3 the distribution of $H$ (top panel) and $S$ (bottom panel), respectively. In each plot, we show separately the distributions of movies with votes in the range $1,000 \leqslant N < 50,000$ (1–50k case) and $N \geqslant 50,000$ (50k+ case); see Supplementary Fig. 2 for the overall distributions. The average $H$ for the 1–50k case is $\bar{H} = 0.45$ and for the 50k+ case is $\bar{H} = 0.40$. The average $S$ for the 1–50k case is $\bar{S} = 0.72$ and for the 50k+ case is $\bar{S} = 0.69$. The latter averages already show that controversy decreases with the number of votes $N$, which we take as an estimator of the "popularity" of a movie. The actual trends of $H$ and $S$ with respect to $N$ are shown in Supplementary Fig. 2.

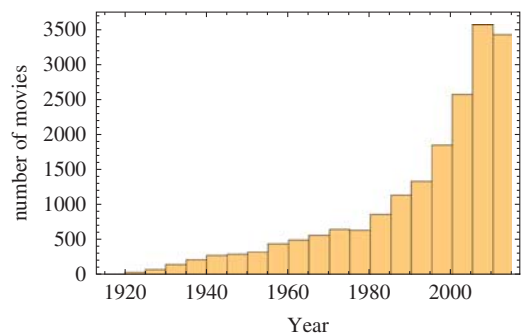

**Figure 1 | Number of movies as a function of film release year.**
**Note:** Only the 19,017 IMDb movies with more than 1000 ratings (as of January 2015) are considered.

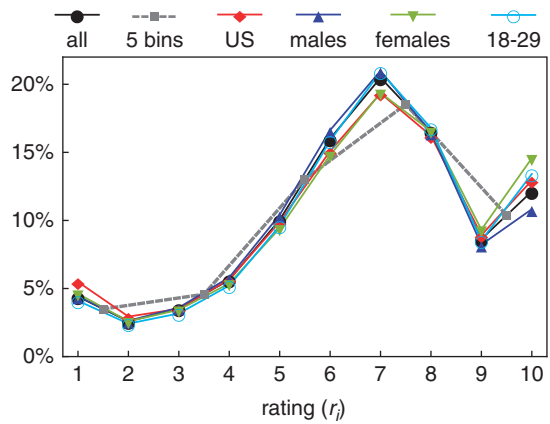

**Figure 2 | Average rating distribution for all movies.**
*Note*: Black: all movies. Dashed grey: all movies, binned according to a 5-star system. Red: only United States of America users. Dark blue: only males. Green: only females. Light blue: only age group 18–29. For the 5-star system, we divide the ratings by 2 and round up, and we also show half the probability for better comparison with the other distributions.

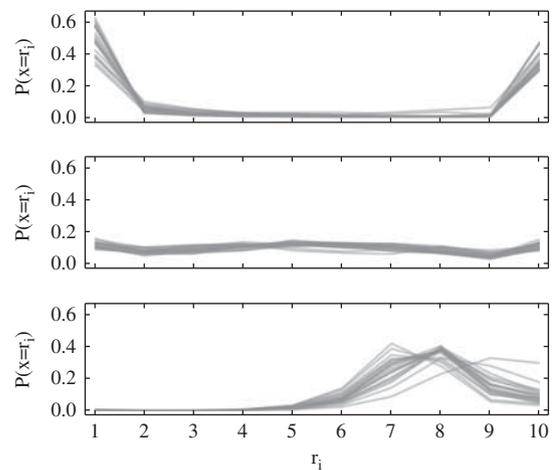

**Figure 4 | High- and low-controversy rating distributions.**
*Note*: Top: Distribution of ratings for the 20 movies with highest hard controversy. In Fig. 5 these movies correspond to the first 20 points from the right. Middle: Same for soft controversy. In Fig. 5 these movies correspond to the first 20 points from the top. Bottom: Same for the 20 least controversial movies (lowest $H$). In Fig. 5 these movies correspond to the first 20 points from the left.

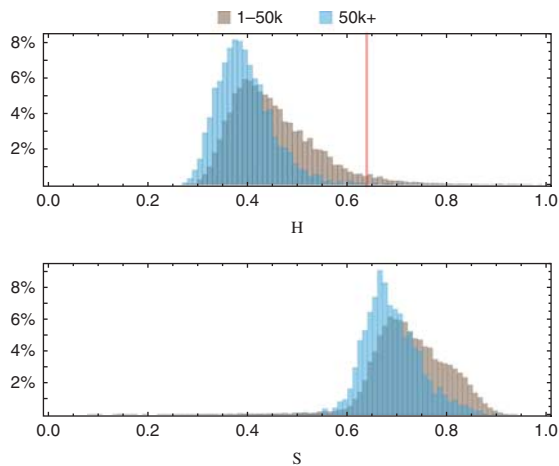

**Figure 3 | Frequency histogram of controversiality.**
*Note*: Top: histogram of frequencies of $H$. The red vertical line marks the threshold of high controversiality $H_{flat}$. Bottom: histogram of frequencies of $S$.

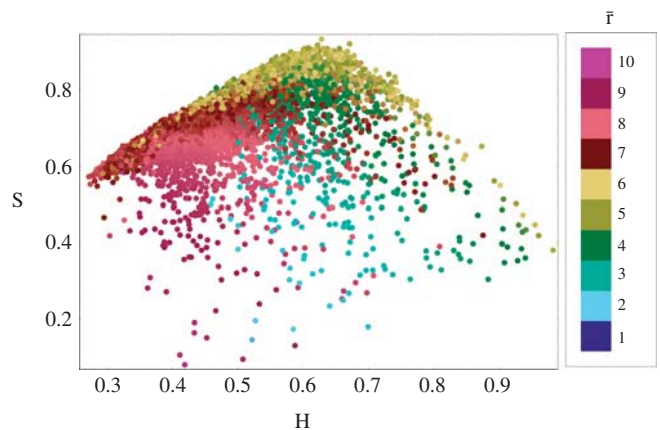

**Figure 5 | Correlation of the hard controversy index $H$ against the soft controversy index $S$.**
*Note*: Movies are represented by dots which are coloured according to their average rating $\bar{r}$.

The distributions of votes for the twenty most controversial movies are shown in Fig. 4 (top and middle panels). As expected, the rating distribution of movies with a high value of $H$ is strongly polarized, while movies with a high value of $S$ have a very flat distribution and all ratings are similarly represented. The distributions of the 20 least controversial movies (lowest $H$) are shown in the bottom panel. Note that these low-controversy movies are not just unanimously bad or good: the votes are instead mostly peaked around 7 or 8, with an approximately Gaussian distribution. Therefore, it appears as if the "concordance vote" is around 7 or 8: people do not seem to be able to agree on, say, a rating of 3. A peak in the ratings around 75% of the highest score has also been reported in Bell and Ippolito (2011) and, on a more episodic basis, in Sun (2012b). We do not only find that both mean and mode of the aggregate rating distribution is around 7–8 (see Fig. 2) but also that movies with this average rating are those with tighter consensus. The correlation between $H$, $S$ and average rating $\bar{r}$ is shown in Fig. 5. Good and bad movies are clustered in adjacent regions. Movies with average rating—those that, as explained below, can have highest $H$ and $S$—are clustered along a boomerang-like region of the parameter space. The fact that this region defines a non-invertible relation between $H$ and $S$ shows the complementarity of the two indexes of controversiality.

A flat distribution may be interpreted as a very broad-peaked distribution. This situation of "no-consensus" corresponds to $H_{flat} = 0.64$ (and of course $S = 1$). Therefore, peaked distributions can only produce controversy indexes in the range $0 \leq H \leq H_{flat}$ and we adopt $H_{flat}$ as the threshold for polarized distributions. The first interesting result we find is that high $H$ controversiality is rare. There are 606 movies (3.6%) with $H > H_{flat}$ for the 1–50k case and only 5 movies (0.24%) for the 50k+ case. To make a comparison, we consider the symmetric value of $H$ around the middle value of 1/2: $H_{low} \equiv 1 - H_{flat} = 0.36$. We find 1,765 movies (10%) with $H < H_{low}$ for the 1–50k case and 605 movies (29%) for the 50k+ case.

It is interesting to note that if one assumes that each rating is composed of a linear combination of independent subcategories (the quality of the actors, screenplay, photography and so on), each with a particular distribution, then the Central Limit

| Table 1 | Number of movies and total number of ratings (in millions) of the various analyses and subcategories of Fig. 6 | | | | | | | |
|---|---|---|---|---|---|---|---|---|
| | | All | The United States | Males | 18–29 | 1–50k | 50k+ | $D \leq 1$ |
| Analysis of $\bar{r}$ | films | 19,017 | 19,017 | 19,017 | 19,017 | 16,924 | 2,093 | 18,855 |
| | ratings | 437M | 93M | 303M | 173M | 123M | 314M | 436M |
| Analyses of $H$ and $S$ | films | 10,540 | 10,753 | 11,035 | 9,927 | 9,651 | 889 | 10,455 |
| | ratings | 161M | 34M | 116M | 55M | 70M | 92M | 161M |

Theorem would imply that the final rating distribution should be well approximated by a Gaussian. Under such an assumption, polarized (high $H$) movies are characterized by a failure of this scenario, with a few subcategories dominating the overall vote or strongly correlated with each other. This issue could be investigated with more detailed data as part of future analysis.

**Release year trends**. As indicated earlier, we show that both $H$ and $S$ decrease with growing $N$. This seems to tell us that either controversial movies do not become very popular or that they lose their controversiality as they do so. The opposite possibility, that controversiality induces a widespread debate that in turn leads to increased popularity (of the kind apparently detected in Sun, 2012a, for low-rated books) is not supported by our findings; perhaps this effect, if present at all, is lost among the other factors of popularity, such as advertisement, language, genre and so on. For more details and a summary plot of the relevant correlations among parameters see Supplementary Information: correlations.

Our database spans a full century of movies and can be further partitioned into various demographic categories according to age group, gender and nationality (see Table 1). Figure 6 shows how the statistics $\bar{r}$, $H$ and $S$ vary with movie release year for the various categories. Several conclusions can be drawn from these results: first, older movies have a substantially higher average rating than more recent movies (see Fig. 6, top panel). A straightforward explanation is that old movies are watched again and rated only if they are good. It is interesting to note—perhaps surprisingly in view of previous studies that showed marked cultural or gender-related differences (Koh et al., 2010; Otterbacher, 2013)—that this trend is robust against age, gender and geographic origin of the voters: the slope does not change significantly when particularizing the analysis to a given demographic subset. The effect of popularity is instead well visible in all plots and implies a marked correlation with $N$ (for more details see Supplementary Information: correlations).

In regard to trends of the controversiality indexes $H$ and $S$, as already mentioned, a movie can have a high $H$ or $S$ only if $\bar{r}$ is not far from 5.5 (otherwise the distribution cannot be polarized nor flat). Therefore, we consider a bin $4 < \bar{r} < 7$ centred around 5.5 in order to remove any correlation of $H$ and $S$ with $\bar{r}$. The trend of $H$ and $S$ shows that controversy increases with release year. This result also remains true after organizing the data according to gender, age group or geographic origin of the votes, and is stronger for popular movies when analysing separately the 1–50k and 50k+ cases. All these trends are shown in the last two panels of Fig. 6.

We also find that, compared to the case "all" of Fig. 6, if one considers separately movies produced before 1990 there is little change in the trends relative to $\bar{r}$ and $H$, and only a moderate increase in slope of the $S$ trend. Therefore, the ever-increasing rate of movie production (see Fig. 1) does not have a significant impact on the release year trends studied in this section.

We note that a source of bias can potentially come from users who try to rig the votes in some way. The IMDb employs a non-disclosed algorithm to minimize the impact of these unrepresentative votes, and the final rating of a movie is obtained from a weighted mean, in which suspicious votes count for less (IMDb, 2015). Therefore, large values of the estimator $D = |$weighted mean $-$ arithmetic mean$|$ should signal movies with rigged rating distributions. We tested the effect of discarding all movies with $D \geq 1$ and concluded it has little impact on the trend of $H$ and $S$ with respect to release year, as can be seen in Fig. 6.

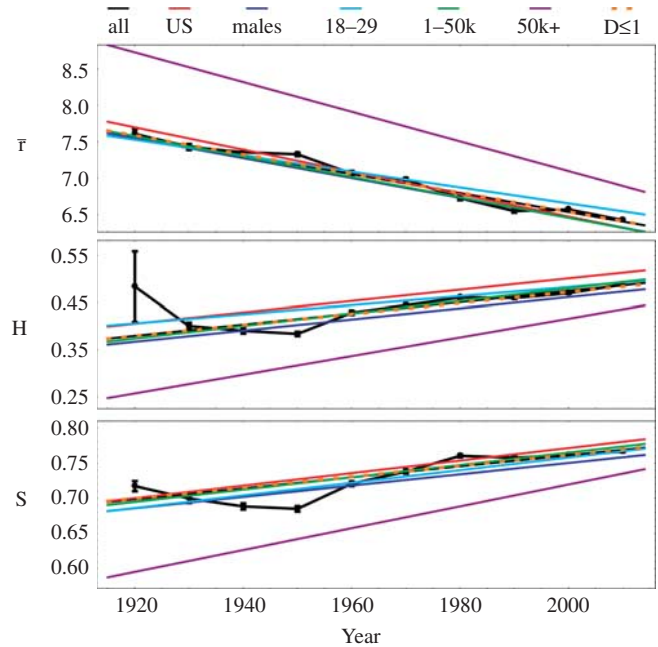

Figure 6 | Release year trends.
Note: Linear regression trends of $\bar{r}$ (top panel), $H$ (middle) and $S$ (bottom) as a function of film release year. In the case of $H$ and $S$ only movies with $4 < \bar{r} < 7$ have been considered. Black: all movies (with votes from all demographics), including error bars representing the error of the mean. Red: only US users. Dark blue: only males. Light blue: only age group 18–29. Green: only movies with less than 50,000 votes. Purple: only movies with more than 50,000 votes. Dotted orange: only movies satisfying the quality cut $D < 1$. See Table 1 for the number of movies and the total number of ratings of the various subcategories.

In summary, we find that the trends of $\bar{r}$, $H$ and $S$ with release year are robust against different binning strategies and against demographic and quality cuts. All the linear regressions shown in Fig. 6 feature a slope that is greater than zero at more than $20\sigma$ confidence level, which corresponds to a $P$-value of $10^{-88}$. The only exception is the case 50k+, which, due to the smaller sample, has a slope greater than zero at $7\sigma$ confidence level ($P$-value of $10^{-12}$). We take our reference sample to be composed of the 9,566 movies with $4 \leq \bar{r} \leq 7$, 1–50k number of votes, quality cut $D \leq 1$ and votes from all demographics together. The corresponding linear regressions are:

$$\bar{r} = 6.5 - 0.014\,(\text{year} - 2000)$$

$$H = 0.48 + 0.0013\,(\text{year} - 2000)$$

$$S = 0.77 + 0.00087\,(\text{year} - 2000) \qquad (4)$$

**Evolution with respect to time of observation.** In the previous section we analysed how average ratings and hard and soft controversiality depend on the movie release year $p$.[3] Here we consider how the parameters $\bar{r}$, $H$ and $S$ depend on the time $t$ of observation, that is, of data collection. In order to perform a robust test of these behaviours, we need two surveys well separated in time. We thus make use of preliminary data collected in January 2013, almost exactly two years before the main data set used in this work. This preliminary data were limited to movies produced between 1950 and 2012 and with more than 5,000 votes, for a total of 6,030 movies, with no breakdown in demographic subgroups.

This analysis is particularly important as far as our subject matter is concerned for it is not obvious how to interpret the fact that old movies appear to be less controversial than recent ones. This might be either because the film industry is producing more controversial movies (the *sophistication* scenario)[4] or because movie-goers are likely to find a film more controversial when it is first produced, but their strong love/hate feelings fade over time (the *convergence* scenario). A combination of both scenarios is of course also possible. This reflects the eternal debate about "modern art" (Gans, 1974; Siegel, 1982; Levine, 1988; Kuspit, 1991; Newman, 2009): is art becoming more and more abstruse and, therefore, controversial, or are people more likely to accept an artistic *avant-garde* when they are temporally distanced from it? In the first scenario, the $H$ index obtained in a survey performed at the time $t$ relative to movies produced in year $p$—we call this quantity $H_p(t)$—remains constant when changing $t$. In the second scenario, $H_p(t)$ decreases when increasing $t$. The same should apply to $S$. As we show below, there is evidence that the *convergence* scenario is favored over the *sophistication* scenario.

In Fig. 7 we show the change of the parameters $\bar{r}$, $H$ and $S$ as a function of the parameter itself for each of the 6,030 movies mentioned above. A clear correlation is evident, which suggests a simple linear model for the average evolution in $t$ of the parameters:

$$\frac{dX_p}{dt} = -\frac{1}{\tau_X}(X_p - X_*) \quad (5)$$

where $X$ represents the value of $\bar{r}$, $H$ and $S$ at time $t$ as due to all votes casted at times previous than $t$. In other words, we are dealing with the temporal evolution of integrated quantities. A positive value of $\tau_X$ means that the evolution converges to $X_*$. The characteristic time $\tau_X$ represents the time scale for such evolution.

A fit of data gives the values listed in Table 2, from which one concludes that hard and soft controversiality of movies do indeed converge to low non-controversial values (compare $H_*$ and $S_*$ with the distributions of Fig. 3) with very similar characteristic times of approximately 40–50 years.[5] After this period, a controversial movie is likely to be classified as fully mainstream. The average rating converges towards 5 with a significantly longer characteristic time. Interestingly, this makes the evolution of $H$ and $S$ become independent of $\bar{r}$ for, as discussed before, the controversiality indexes are uncorrelated with the average rating if the latter takes central values. The statistical significance of the characteristic times is always very high, as one can infer from the errors quoted in Table 2.

Three important temporal factors could affect the voting trends in recent decades. First, movies released more recently have had a shorter voting time than earlier movies. Furthermore, since IMDb launched in 1990, only movies released after this date had a chance of being voted for when first screened, when one expects movies to be discussed more—except for famous classic films. Finally, internet usage has increased dramatically and continuously since IMDb was launched. Within our data this is demonstrated by the fact that the average number of ratings has grown significantly with respect to the release year (see Supplementary Information: Correlations). In order to study how important the consequences of these trends are, we show in Table 2 the characteristic times for movies produced before and after 1990. We obtain similar parameters regarding $S$ and $\bar{r}$ but 20% different characteristic times as far as $H$ is concerned. A more thorough assessment of these effects would involve analysing the evolution of non-integrated estimators. This can form the basis of future research.

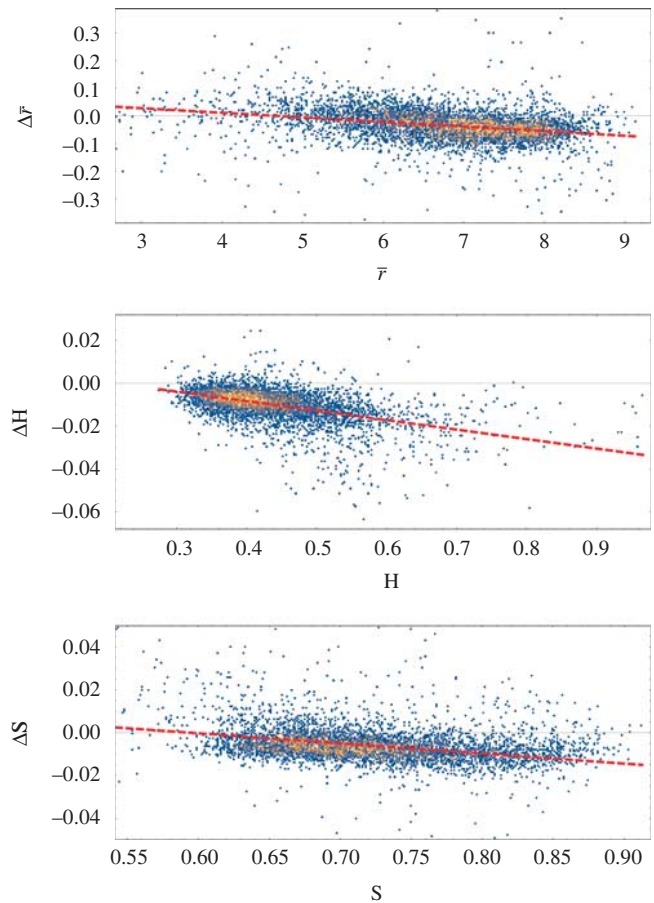

**Figure 7 | Evolution of average rating and controversiality.**
*Note*: Change of $\bar{r}$, $H$ and $S$ between 2013 and 2015 (2 years) as a function of $\bar{r}$, $H$ and $S$, respectively, for the 6,030 movies produced between 1950 and 2012 with more than 5,000 ratings. The red dashed line represents a linear regression to the data. The bins in the plots are color coded blue-to-yellow according to how many movies they contain.

Then we studied how the characteristic times depend on the number of votes $N$. Table 2 shows the characteristic times of $\bar{r}$, $H$ and $S$ for two non-overlapping bins in popularity and Fig. 8 shows the characteristic times for movies with number of votes greater than a given number in the $x$-axis. A clear trend is evident: the controversiality of popular movies decreases faster than that of less popular movies. This is in agreement with the findings in Fig. 6, where very popular movies are seen to show a faster rise in controversiality with release year then less popular ones. This could imply that movies that are more "discussed" reach consensus more rapidly, that is, converge faster to a peaked distribution. This seems to support a scenario of "convergence through interaction", as often studied in social dynamics (Baldassarri and Bearman, 2007; Castellano *et al.*, 2009), in contrast to the opposite pattern of the so-called "echo chamber" effect (Jamieson and Cappella, 2010), that is, strong interaction only within like-minded communities. The evolution of soft controversy seems to be more sensitive to popularity than hard controversy. Finally, the same analysis relative to $\bar{r}$ shows that the characteristic time increases, suggesting that popular movies have a more stable average rating.

Table 2 | Fits of the parameters of equation (5) for all movies, and for two bins in popularity and release year

|  | all | $5 \cdot 10^3 \leq N < 5 \cdot 10^4$ | $5 \cdot 10^4 \leq N < 10^6$ | $1950 \leq yr < 1990$ | $1990 \leq yr < 2013$ |
|---|---|---|---|---|---|
| films | 6,030 | 4,627 | 1,403 | 1,439 | 4,591 |
| $\tau_{\bar{r}}$ | 120 ± 5 | 113 ± 5 | 170 ± 20 | 110 ± 7 | 109 ± 5 |
| $\bar{r}_*$ | 4.7 ± 0.5 | 4.8 ± 0.5 | 4 ± 1 | 5.6 ± 0.8 | 4.7 ± 0.5 |
| $\tau_H$ | 45 ± 1 | 42 ± 1 | 41 ± 2 | 37 ± 2 | 48 ± 1 |
| $H_*$ | 0.21 ± 0.02 | 0.24 ± 0.02 | 0.19 ± 0.03 | 0.25 ± 0.03 | 0.20 ± 0.02 |
| $\tau_S$ | 42 ± 2 | 42 ± 2 | 33 ± 2 | 41 ± 2 | 43 ± 2 |
| $S_*$ | 0.60 ± 0.05 | 0.61 ± 0.05 | 0.59 ± 0.07 | 0.60 ± 0.07 | 0.59 ± 0.06 |

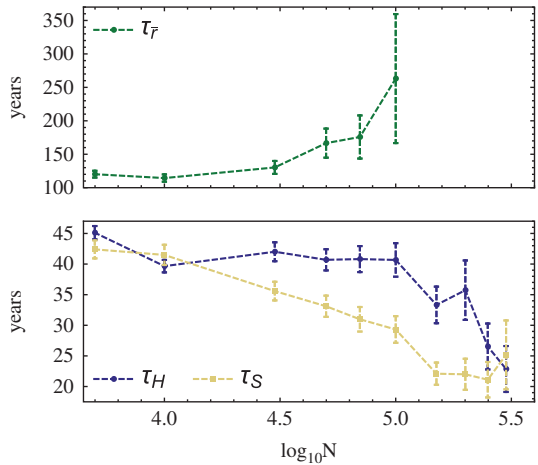

**Figure 8 | Dependence of characteristic time on number of votes N.**
*Note*: Characteristic times for the set of movies with number of votes greater than the number in the *x*-axis. Note that for large value of *N* the slope of the correlation relative to $\bar{r}$ shown in Fig. 7 is consistent with zero; therefore, the corresponding very large values of the characteristic times are not shown in the plot.

As we have seen, the fitted parameters depend on the movie popularity, which in turn depends on *t*. Nonetheless, as the dependence is not very strong we can treat $\tau_X$ as constant and integrate the previous equation to obtain the following approximate exponential behaviour:

$$X_p(t) - X_* = (X_0 - X_*) e^{-\frac{t-t_0}{\tau_H}} \quad (6)$$

where $X_0 = X_p(t_0)$.

The results outlined here support the idea that while all movies are created on average equal (i.e. the distribution of the initial amount of controversiality is almost independent of release year), the subsequent evolution de-polarizes the audiences. As a consequence, today older movies appear less controversial than modern ones. It is interesting to note that previous works on online rating trends reported either an increase in mean and a decrease in variance (Zhang *et al.*, 2014), or vice versa (Moe and Schweidel, 2012). Our findings, that the mean converges to the central value of the spectrum and at the same time the variance decreases (though with a different characteristic time), a phenomenon that we term "convergence to mainstream", appear to be novel.

### Future work
One of our main results is that controversial films slowly lose their controversiality as time passes, converging to mainstream in the scale of 4–5 decades in a way described by an exponential decay. Although this simple model fits the data well, it might be the case that the true governing law is different. Since we have only used two datasets, collected in January 2013 and January 2015, we cannot elaborate further on this issue. One would need data collected at further epochs to get a more refined understanding of the convergence. Another limitation is that since our 2013 survey did not include demographic data we cannot estimate the dependence of the characteristic times on different user categories.

Further future research could involve obtaining data that includes the epoch of each vote. This would allow us to study the evolution of instantaneous, rather than integrated, parameters (*H*, *S* and $\bar{r}$), which would prove helpful in order to understand the underlining dynamics of patterns of controversiality in movies.

Our results, like all those based on online ratings or reviews, may also be limited by underreporting biases (Hu *et al.*, 2009) and other unknown biases inherent in every survey (Schuman and Scott, 1987). It must be clear that here we can only discuss the controversiality among the IMDb raters and not of the general public at large; still, invoking the same argument as in Schuman and Scott (1987), we might expect that the *changes*, rather than the absolute values, within the same group are a robust indicator of public opinion.

As Koh *et al.* (2010) identified, US users in IMDb strongly underreport their opinions, distinctively more so than Chinese users of douban.com. Since we find no strong difference among US users and non-US users in our IMDb data, this may imply that all IMDb users have similar behaviour (which in itself merits further investigation). Koh *et al.* (2010) also claim that other websites based on a recommendation system (like MovieLens) and not purely on online reviewing are less subject to underreporting bias. Therefore, it would be valuable to conduct a similar study on data from other movie review sites.

### Conclusions
In the present work we address the issue of controversiality of films from a quantitative and statistical point of view. We find that old movies are significantly better rated than more recent ones, and popular movies are significantly better rated than less popular ones. In addition, we find that controversy decreases with popularity and that hard controversy is relatively rare: only 3.6% of the movies are above the value that separates peaked distributions from polarized distributions. Furthermore, we find that modern movies are judged to be more controversial than old ones and, at the same time, we find a trend of "convergence to mainstream" with a time scale of roughly 40–50 years. A decrease in the characteristic time scale with popularity seems to support a scenario of convergence through interaction. We also find that the average rating converges to the central value of the spectrum with a longer time scale of 100–150 years. These results have been found to be robust with respect to demographic breakdown (whenever we could perform such breakdown). Finally, we observe that the overall distribution of votes does not follow a simple $J$ or $U$-shape (see Supplementary Information: the average rating distribution), as other online review systems do (Hu et al., 2009; Koh et al., 2010). Instead, it exhibits a double edge-peaked trimodal distribution, resembling a $J^V$ shape, which is also universal among demographic subgroups. The causes behind this unusual shape merit further analysis.

The history of art, and of culture in general, has witnessed the phenomenon in which avant-gardes that have been initially considered controversial, have then over time become accepted as part of the mainstream (unless they disappear without leaving significant traces). It is worth noting the many derisive epithets initially employed against avant-gardes (for example, impressionism, fauvism, constructivism, cubism) that are today the mainstay of the most important museums of modern art. Similar processes seem also to constellate the scientific debate, where consensus ultimately often emerges after years or decades of heated discussion, as in the famous controversy over the foundations of quantum mechanics between, among others, Einstein and Bohr, or the debate of early twentieth century astronomy on the nature of nebulae.[6] Our results, although limited to a particular form of expression, movies, might be seen to be a confirmation of these phenomena. In contrast, political or ethical debate is often characterized by increasing polarization (or ideological radicalization, as is often denoted in this context) (Sunstein, 2002; Garcia et al., 2015; Koutra et al., 2015), although the extent of this process has been debated in, for instance, DiMaggio et al. (1996), Fiorina and Abrams (2008) and Fischer and Mattson (2009). When radicalization occurs, sometimes it has been attributed to the "echo chamber" effect, a reinforcing of previous opinion through interaction only within close and homogeneous communities. One can then hypothesise that echo chambers are harder to build or to maintain in the artistic or cultural realm than in the political or ethical one, where organized groups might have a great and direct advantage in polarizing their audiences. In view of a broader discussion of such speculative arguments, the extension of the quantitative analysis of controversiality to other online databases with different demographic user bases and other forms of cultural expression such as musical and literary compositions appears to be a pressing task.

## Notes

1 Similar controversy patterns are found in other areas tarrow (Dascal, 1998; Tarrow, 2008).
2 This threshold refers to the full dataset; demographic subgroups can have fewer votes.
3 In the previous sections we used "year" to denote $p$. We will use "year" and $p$ interchangeably.
4 As pointed out by Douglas Adams in *The Hitchhiker's Guide to the Galaxy: The Restaurant at the End of the Universe*, the third phase of every major galactic civilization is sophistication.
5 "Au théâtre, comme dans tous les arts, il y a les gens qui voient et les aveugles-nés. Il faut toute une vie à ces derniers pour s'habituer aux grandes choses et ce n'est qu'aprés avoir entendu rabâcher pendant cinquante ans: ceci est beau, qu'ils se rendent au jugement des autres". A 1923 quote from French theatre director charles ullin, cited in Bishop (1964).
6 Many other scientific case-studies are discussed in, for example, Machamer et al. (2000).

## References

Akoglu L (2014) Quantifying political polarity based on bipartite opinion networks. In *International AAAI Conference on Web and Social Media*, North America, May, http://www.aaai.org/ocs/index.php/ICWSM/ICWSM14/paper/view/8073.

Baldassarri D and Bearman P (2007) Dynamics of political polarization. *American Sociological Review*; **72** (5): 784–811.

Bell J and Ippolito J (2011) When the rich don't get richer: Equalizing tendencies of creative networks. *Leonardo*, **44** (3): 260–261.

Bishop T (1964) Changing concepts of avant-garde in xxth century literature. *The French Review*; **38** (1): 34–41.

Cai T, Cai H J, Zhang Y, Huang K and Xu Z (2013) Polarized score distributions in music ratings and the emergence of popular artists. In *Science and Information Conference (SAI)*, 7–9 October, pp 472–476, http://ieeexplore.ieee.org/xpl/abstractAuthors.jsp?arnumber=6661781.

Castellano C, Fortunato S and Loreto V (2009) Statistical physics of social dynamics. *Reviews of Modern Physics*; **81** (2): 591–646.

Conover M D, Ratkiewicz J, Francisco M, Goncalves B, Menczer F and Flammini A (2011) Political polarization on twitter. In *International AAAI Conference on Web and Social Media*, North America, July, https://www.aaai.org/ocs/index.php/ICWSM/ICWSM11/paper/view/2847.

Dascal M (1998) Types of polemics and types of polemical moves. In Cmejrkova JH, Mullerova O and Svetla J (eds) *'S', Max Niemeyer, Dialogue Analysis VI (Proceedings of the 6th Conference, Prague 1996)*, Vol. 1. Tubingen, pp 15–33.

DiMaggio P, Evans J and Bryson B (1996) Have american's social attitudes become more polarized? *American Journal of Sociology*; **102** (3): 690–755.

Fiorina M P and Abrams S J (2008) Political polarization in the american public. *Annual Review of Political Science*; **11** (1): 563–588.

Fischer C S and Mattson G (2009) Is America fragmenting? *Annual Review of Sociology*, **35**, 435–455.

Gans H (1974) *Popular Culture and High Culture. An Analysis and Evaluation of Taste*. Basic Books: New York.

Garcia D, Abisheva A, Schweighofer S, Serdült U and Schweitzer F (2015) Ideological and temporal components of network polarization in online political participatory media. *Policy & Internet*; **7** (1): 46–79.

Garimella K, De Francisci Morales G, Gionis A and Mathioudakis M (2015) Quantifying controversy in social media. *ArXiv e-prints 1507.05224*, http://arxiv.org/abs/1507.05224.

Godes D and Silva J C (2012) Sequential and temporal dynamics of online opinion. *Marketing Science*; **31** (3): 448–473.

Hofstede G (2001) *Culture's Consequences: Comparing Values, Behaviors, Institutions, and Organizations Across Nations*, second edition, Sage Publications: CA.

Holbrook M (2005) The role of ordinary evaluations in the market for popular culture: Do consumers have "good taste"? *Marketing Letters*; **16** (2): 75–86.

Hoßfeld T, Schatz R and Egger S (2011) SOS: The MOS is not enough! In *Quality of Multimedia Experience (QoMEX), 2011 Third International Workshop*, IEEE, 7–9 September, pp 131–136, http://ieeexplore.ieee.org/xpl/abstractAuthors.jsp?arnumber=6065690.

Hu N, Zhang J and Pavlou P A (2009) Overcoming the j-shaped distribution of product reviews. *Communications of the ACM*; **52** (10): 144–147.

IMDb. (2015) http://www.imdb.com/help/showleaf?votes.

Jamieson K H and Cappella J N (2010) *Echo Chamber. Rush Limbaugh and the Conservative Media Establishment*. Oxford University Press: New York.

Koh N S, Hu N and Clemons E K (2010) Do online reviews reflect a product's true perceived quality? An investigation of online movie reviews across cultures. *Electronic Commerce Research and Applications*; **9** (5): 374–385, Special Section on Strategy, Economics and Electronic Commerce.

Kostakos V (2009) Is the crowd's wisdom biased? A quantitative analysis of three online communities. In *Computational Science and Engineering. CSE '09. International Conference*, Vol. 4, 29–31 August 2009, pp 251–255.

Kostelanetz R (2000) *A Dictionary of the Avant-Gardes*. Psychology Press: Routledge, London.

Koutra D, Bennett P N and Horvitz E (2015) Events and controversies: Influences of a shocking news event on information seeking. In *Proceedings of the 24th International Conference on World Wide Web (WWW)*. International World Wide Web Conferences Steering Committee, Republic and Canton of Geneva, Switzerland, pp 614–624, http://dl.acm.org/citation.cfm?id=2736277.2741099.

Kuhn T S (1962) *The Structure of Scientific Revolutions*. University of Chicago Press: Chicago.

Kuspit D (1991) The appropriation of marginal art in the 1980s. *American Art*; **5** (1): 132–141.

Levine L (1988) *Highbrow/lowbrow: The Emergence of Cultural Hierarchy in America*. Harvard University Press: Cambridge, MA.

Liu Y (2006) Word of mouth for movies: Its dynamics and impact on box office revenue. *Journal of Marketing*; **70** (3): 74–89.

Machamer P, Pera M and Baltas A (2000) *Scientific Controversies: Philosophical and Historical Perspectives*. 1st edn. Oxford University Press: New York.

McCarty N, Poole K T and Rosenthal H (2006) *Polarized America: The Dance of Ideology and Unequal Riches*, Walras-Pareto Lectures, MIT Press.

Mejova Y, Zhang A X, Diakopoulos N and Castillo C (2014) Controversy and sentiment in online news. Computation and Journalism Symposium 2014. 24–25 October. Columbia University, New York, NY.

Moe W W and Schweidel D A (2012) Online product opinions: Incidence, evaluation, and evolution. *Marketing Science*; **31** (3): 372–386.

Moon S, Bergey P K and Iacobucci D (2010) Dynamic effects among movie ratings, movie revenues, and viewer satisfaction. *Journal of Marketing*; **74** (1): 108–121.

Newman M Z (2009) Indie culture: In pursuit of the authentic autonomous alternative. *Cinema Journal*; **48** (3): 16–34.

Otterbacher J (2013) Gender, writing and ranking in review forums: A case study of the IMDb. *Knowledge and Information Systems*; **35** (3): 645–664.

Pennacchiotti M and Popescu A-M (2010) Detecting controversies in twitter: A first study. In *Proceedings of the NAACL HLT 2010 Workshop on Computational Linguistics in a World of Social Media*, WSA '10, Association for Computational Linguistics, Stroudsburg, PA, pp 31–32, http://dl.acm.org/citation.cfm?id=1860667.1860683.

Rad H S and Barbosa D (2012) Identifying controversial articles in Wikipedia: A comparative study. In *Proceedings of the Eighth International Symposium on Wikis and Open Collaboration*, 27–29 August. WikiSym: Linz, Austria.

Schuman H and Scott J (1987) Problems in the use of survey questions to measure public opinion. *Science*; **236** (4804): 957–959.

Siegel M B (1982) Vanguard meets the mainstream. *The Hudson Review*; **35** (1): 99–104.

Sun M (2012a) How does the variance of product ratings matter? *Management Science*; **58** (4): 696–707.

Sun S (2012b) *Why reader review metascores are meaningless*, http://stephsun.com/metascores.html, accessed November 2015.

Sunstein C R (2002) The law of group polarization. *Journal of Political Philosophy*; **10** (2): 175–195.

Tarrow S (2008) Polarization and convergence in academic controversies. *Theory and Society*; **37** (6): 513–536.

Wanderer J J (1970) In defense of popular taste: Film ratings among professionals and lay audiences. *American Journal of Sociology*; **76** (2): 262–272.

Yardi S and Boyd D (2010) Dynamic debates: An analysis of group polarization over time on twitter. *Bulletin of Science, Technology & Society*; **30** (5): 316–327.

Yasseri T, Spoerri A, Graham M and Kertész J (2014) The most controversial topics in wikipedia: A multilingual and geographical analysis. In: Fichman P and Hara N (eds) *Global Wikipedia: International and Cross-Cultural Issues in Online Collaboration*. Rowman & Littlefield Education, pp 25–48.

Zhang Y, Lappas T, Crovella M and Kolaczyk E D (2014) Online ratings: Convergence towards a positive perspective? In *Acoustics, Speech and Signal Processing (ICASSP)*, 2014 IEEE International Conference, 4–9 May, pp 4788–4792, http://ieeexplore.ieee.org/xpl/articleDetails.jsp?arnumber=6854511.

## Data Availability

The datasets analysed during the current study are not publicly available due IMDb conditions of use but are available from the corresponding author on reasonable request.

## Author Contributions

Luca Amendola, Valerio Marra and Miguel Quartin contributed equally to this work.

## Acknowledgements

Datasets were courtesy of IMDb (http://www.imdb.com) and were used with permission. The authors have no affiliation with IMDb; nor is IMDb involved any way in our study. We thank Fabiana V. Campos, Giulio Marra and Ana Luisa Santos for useful comments. In Fig. 5 we have adopted the color palette optimized for colour-blind people developed by Paul Tol (available at personal.sron.nl/pault). MQ is grateful to Brazilian research agency CNPq for support and to the University of Heidelberg for hospitality.

## Additional Information

**Supplementary Information:** accompanies this paper at http://www.palgrave-journals.com/palcomms

**Competing interests:** The authors declare no competing financial interests.

**Reprints and permission** information is available at http://www.palgrave-journals.com/pal/authors/rights_and_permissions.html

**How to cite this article:** Amendola L, Marra V and Quartin M (2015) The evolving perception of controversial movies. *Palgrave Communications*. 1:15038 doi: 10.1057/palcomms.2015.38.